\documentclass{llncs}

\input{psfig.sty}
\begin{document}
\pagestyle{headings}
\frontmatter

\title{Divergent evolution paths of different genetic families in the Penna
model}
\author{Miko{\l}aj Sitarz \and Andrzej Maksymowicz}

\institute{Faculty of Physics and Applied Computer Science\\
AGH-University of Science and Technology\\
Al. Mickiewicza 30\\
30-059 Krak{\'o}w, Poland\\
\email{sitarz@novell.ftj.agh.edu.pl}}

\bibliographystyle{splncs}

\maketitle
\begin{abstract}
We present some results of simulations of population growth and evolution, using
the standard asexual Penna model, with individuals characterized by a string of
bits representing a genome containing some possible mutations.  After about
$20000$ simulation steps, when only a few genetic families are still present
from among rich variety of families at the beginning of the simulation game,
strong peaks in mutation distribution functions are observed.  This known effect
is due to evolution rules with hereditary mechanism.  The birth and death
balance in the simulation game also leads to elimination of families specified
by different genomes. Number of families $G(t)$ versus time $t$ follow the power
law, $G \propto t^n$. Our results show the power coefficient exponent $n$
is changing as the time goes.  Starting from about --1, smoothly achieves about --2 after
hundreds of steps, and finally has semi-smooth transition to 0, when only one
family exists in the environment.
This is in contrast with constant $n$ about --1 as found, for example, in
\cite{bib:evolution}. We suspect that this discrepancy may be due to two
different time scales in simulations - initial stages follow the $n\approx-1$
law, yet for large number of simulation steps we get $n\approx-2$, providing
random initial population was sufficiently big to allow for still reliable
statistical analysis.  The $n\approx-1$ evolution stage seems to be associated
with the Verhulst mechanism of population elimination due to the limited
environmental capacity
 - when the standard evolution rules were modified, we observed a plateau
 ($n=0$) in the power law in short time scale, again followed by $n\approx -2$
 law for longer times. The modified model uses birth rate controlled by the
 current population instead of the standard Verhulst death factor.
\end{abstract}

\section*{Introduction}
The Penna model \cite{bib:penna1} is a one of variety of mutation accumulation
models. They are based on assumption that biological ageing is caused by
deleterious mutations in genome (it is inherited from parent -- asexual model --
and enriched by additional mutations put at random site at genome at the birth
moment).

The 
number of genetic families in the population decreases as the time grows. This
was described for example in papers \cite{bib:penna1}, \cite{bib:eveeffect}.
The effect of the reduction in number of genetic families is also responsible
for sharp and irregular maxima in the mutation distribution function after many
time steps of computer simulation. This effect is discussed in the main text.

\section*{Model}
In the model presented, individuals genome is represented as a string of bits,
where '$1$' means presence of mutation, and '$0$' -- lack of mutation. When
an individual reaches age $a$, only the first $a$ bits are exposed. The number of
``ones'' in that part of genome are considered as active mutations $m$. When the
number $m$ of active mutations exceeds a threshold $T$, $m \geq T$ (one of the
model parameters) -- the individual dies due to genetic reason.  Another important
parameter is $R$ -- minimal reproduction age -- only individuals with $a \geq R$
can give offspring. That causes some genomes not to be reproduced, because
its owner dies before reaching mature age. This is the basic
implementation of Darwinian evolution mechanism.

There is also another reason of death incorporated into the model -- death due
to overpopulation -- controlled by so called Verhulst factor. The probability of
such death equals to $n(t)/N$, where $n(t)$ is number of individuals in
time step $t$, and $N$ is the environmental capacity (which is constant during
the simulation). This factor limits the population size.

We start the simulation with population which has no mutations (genome length
is $32$ bits). It takes several hours using Pentium III class machine.  In each
time step:
\begin{itemize}
    \item each individual is tested against possible death, due to:
	\begin{itemize}
	    \item genetic reason, when $m \geq T$;
	    \item Verhulst factor with probability $n(t)/N$;
	\end{itemize}
    \item if individual survived and reached age $R$, it gives birth
	to $B$ babies on average, so $B$ is the birth rate.
	Each descendant gets the genome inherited from parent, enriched by $M$ new
	mutations.
        (If one of the $M$ mutation attempts hits an already mutated bit,
        this bit stays as it is.)
    \item all survivals and newly born individuals make up population at the
	next time step.
\end{itemize}

\section*{Population structure}
By a population structure we mean set of parameters, we use to
characterize it. The first one is the survival factor:
\begin{equation}
    s(a,t) = \frac{n(a+1,t+1)}{n(a,t)},
\end{equation}
where $n(a,t)$ is number of individuals of age $a$ at time $t$. This survival factor is
interpreted as a probability that individual which has age $a$, enters the next
time step $t+1$. From this -- we can obtain mortality rate:
\begin{equation}
    \label{eq:mort}
    q(a,t) = 1 - s(a,t) = 1 - \frac{n(a+1,t+1)}{n(a,t)}
\end{equation}

\begin{figure}
\centerline{\psfig{figure=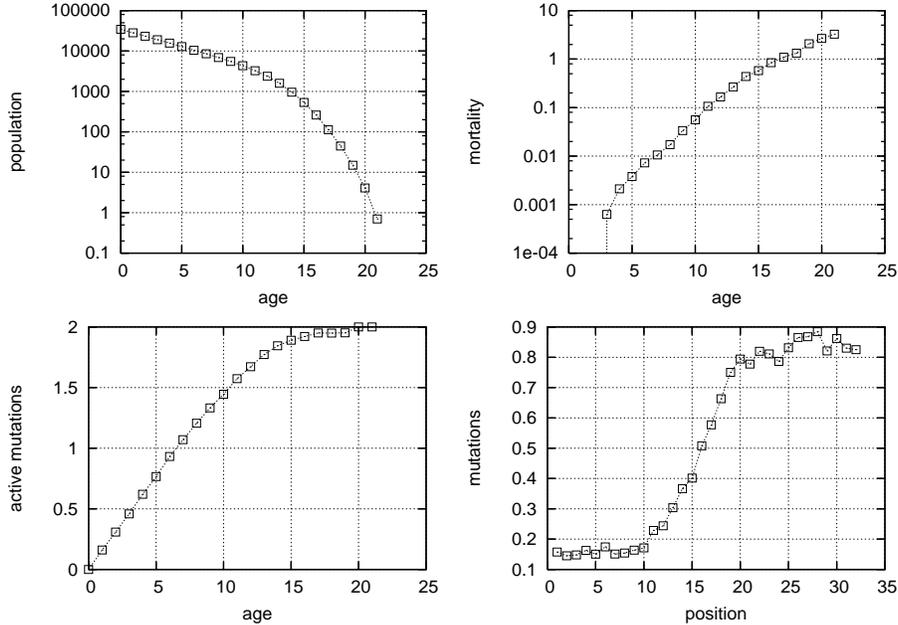,width=1.0\textwidth}}
\caption{Population structure after $1000$ steps. Standard Penna model
$(N,B,R,M,T)=(10^6,1.0,7,1.0,3)$.}
\label{fig:struct1000}
\end{figure}

We can separate mortality for two elements -- mortality because of Verhulst
factor, and mortality because of achieving threshold $T$ by number of active mutations:
\begin{equation}
    \label{eq:mortsep}
    q(a,t) = q_v(a,t) + q_g(v,t) = \frac{v(a,t) + g(a,t)}{n(a,t)},
\end{equation}
where $v(a,t)$ is number of individuals killed because of Verhulst factor, and
$g(a,t)$ is number of deaths caused by genetic reason. The 
quantity $q_v(a,t) = q_v(t)$ is age-independent, whereas $q_g(a) = \lim_{t \to \infty} q_g(a,t)$
gives approximately dependency:
\begin{equation}
    \label{eq:gompertz}
    q_g(a) \approx q_0 \cdot \exp(a \cdot b),
\end{equation}
(where $q_0$ and $b$ are constants) which is known as Gompertz law
\cite{bib:evolution}. It is an empirical law observed also in human population (see
\cite{bib:popreference}, \cite{bib:mortorg}).

Figure \ref{fig:struct1000} presents results after $1000$ time-steps of
simulation (averages from last 20 steps), when the average number of births equals
average numbers of deaths. As can be seen at the chart in the right top corner --
Gompertz law is fulfilled (it presents $q_g(a)$). The chart at the right down corner,
presents the average number of mutations versus bit position. Low values at the
beginning are the effect of evolution mechanism, that eliminates genotypes with
big number of mutations at this part of genome (individuals with such genome can
not reproduce because they die before reaching age $R$). As it can be seen in
the next section, at this case the population is quite diverse and the result
significantly differs from that after $20000$ steps (the same parameters,
figure \ref{fig:struct20000}).

\section*{Decline of a number of genetic families} \label{sec:decline}
The population characteristics presented in the previous section (after $1000$
steps) could lead us to misleading conclusions. Average number of birth acts and
average number of deaths are equals (and what follows -- average population
size does not change), so it may look like a stationary state. 
Indeed, it is so in terms of the population size, however the evolution is still
shifting the genome distribution function.
The proportions of participation of particular genotypes in population are changing.
It is so since first we need to define the criterion with respect to which the
close-to-stationary state is claimed. If the total population $n(t)$ is
considered, one hundred iterations may be enough. However if we mean overall
characteristics of genomes, it is far from the equilibrium and we must proceed
simulation for many thousands steps. There are different time scales for
different features observed in population.

To distinguish between different genotypes -- we introduced a special marker of
individuals. At the beginning of simulation -- each individual inserted into
environment, has given unique number. When individual gives birth -- the marker
is inherited by progeny. The marker is kept fixed which makes it possible to trace
the evolution of different genetic families (by genetic family we mean set of
individuals with the same marker).

\begin{figure}
\centerline{\psfig{figure=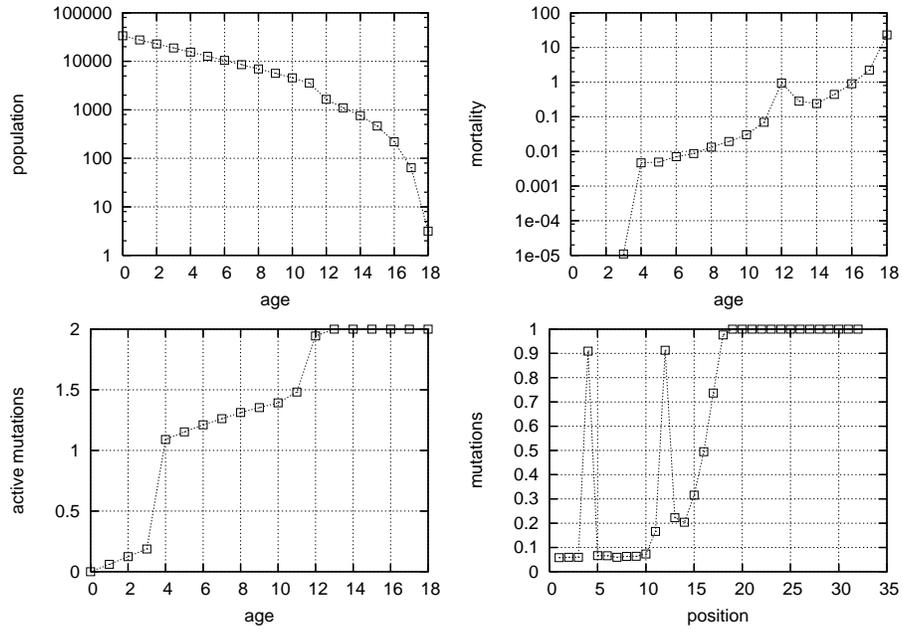,width=1.0\textwidth}}
\caption{Population structure after $20000$ steps. Standard Penna model
$(N,B,R,M,T)=(10^6,1.0,7,1.0,3)$. Strong peaks in mutations distribution
(result of Eve effect) is also reflected in the mortality $q(a)$ case.}
\label{fig:struct20000}
\end{figure}

\begin{figure}
\centerline{\psfig{figure=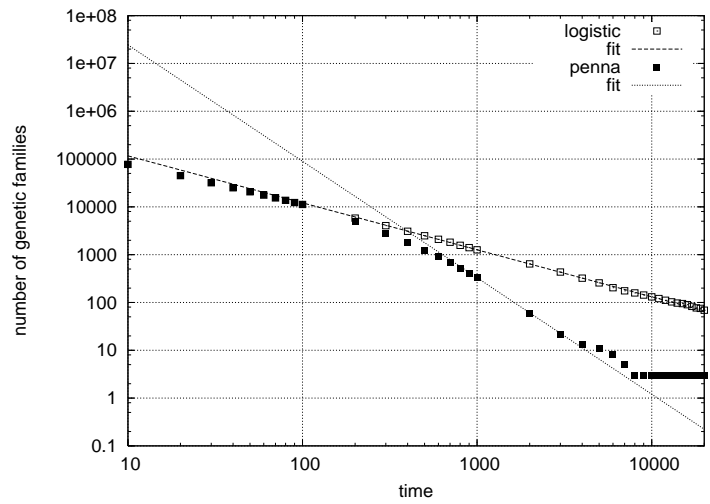,width=0.8\textwidth}}
\caption{Number of genetic families in environment as a function of time. Full
squares -- Penna model, empty squares -- logistic model.}
\label{fig:glines}
\end{figure}

Figure \ref{fig:struct20000} presents results of simulation after $20000$ steps
(this is the continuation of simulation presented at the previous section).
Peaks in mutations distribution -- the right down corner -- suggest that the population
became more homogeneous. To check our hypothesis we plotted the number $G(t)$ of genetic
families versus the time step -- full squares at figure \ref{fig:glines}
(double log scale).  As can be seen it diminishes rapidly -- and after about
$500$ steps becomes a power function of time.

\begin{figure}
\centerline{\psfig{figure=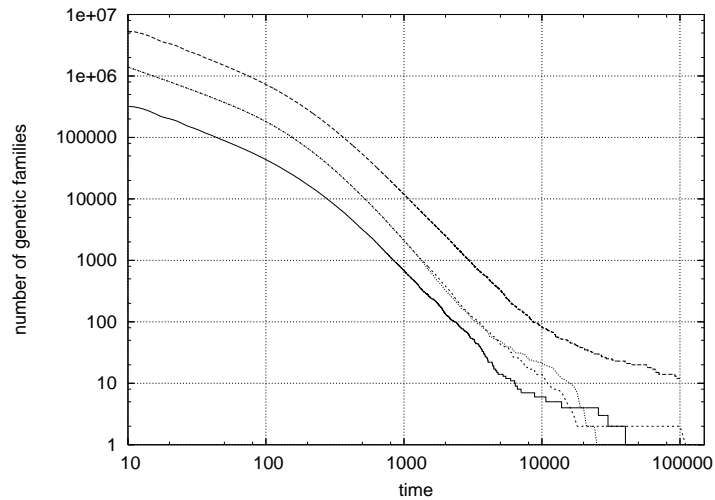,width=0.8\textwidth}}
\caption{Number of genetic families versus time -- four different
simulations. The most top curve (dashed) and the most bottom (solid) are
obtained from results given us by Dietrich Stauffer \cite{bib:dstauffer} 
(for initial population equals accordingly $10^7$ and $6\cdot10^5$).  Two
curves in the middle (dotted and short dashed) originate from our simulations --
all parameters are the same in this two cases except pseudo-random number
generator initial value (initial population equals $2 \cdot 10^6$).}
\label{fig:bigsim}
\end{figure}

In our case -- starting with as many as $4\cdot10^5$ genetic families, after
about $20000$ steps we got three families in environment. For a long time run we
get results presented in figure \ref{fig:bigsim}.

As it was mentioned earlier, we observe that $G(t)$ dependency is:
\begin{equation}
    G(t) \sim t^n
\end{equation}
It seems to be independent on Penna model parameters as long as we not approach
to the logistic model -- empty squares at figure \ref{fig:glines} (by logistic
model, we mean model where the only reason of death is overpopulation, no genetic
factor is present).  This result is compatible with results described in
\cite{bib:evolution} but with one exception. In our simulations $n \approx -2$
for asexual model ($n \approx -1$ for logistic one), while in 
literature $n \approx -1$ is claimed.
We suppose that this discrepancy is caused by the fact, that we start simulation
with much bigger initial population, and so we have had a chance to achieve the
$n \approx -2$ region, that is before the population is extinct.

Our three families as the final result, is rather an effect of under-counting
the simulation. If the simulation time is sufficiently big, the number of
families achieves $1$ every time  -- see figure \ref{fig:bigsim}.  This
phenomena is known as ``Eve effect''.  What is more, the $G(t)$ curve achieves
$n \approx -1$ region again at the end.  So there are three stages of that play:
$-1$ from the start, to several hundred time-steps, then $-2$ until time reaches
about $10000$ steps, and final $-1$ (figure \ref{fig:bigsim}).

We would like to emphasize, that whole ``s-shape'' effect does not occur in
sexual model. In that case -- whole curve is without any transitions, and has an
exponent equals $-1$ (what remains in accordance with results presented in
\cite{bib:evolution}).

We suppose that an important role in the presented ``Eve effect'' plays the
Verhulst factor.  This is the only way which one family can influence on
another. When one family accidently outnumbers another -- it can produce more
new individuals filling up the environment, and death because of overpopulation
makes not as big damage as in less numerous families.  Interesting could be the
introduction of some mechanism into the model, that allows for exchanges between
families which may change the final result.

\section*{Modified Penna model - with no Verhulst factor}
Different mechanisms leading to elimination of individuals gives insight to
the evolution game. The Verhulst factor, however, was introduced in the past
as a remedy to avoid unlimited growth of population. It was not devised to
account for specific death cause such as disease, genetic death, hunting etc.
When we go for more detailed modeling, one should replace the overall
Verhulst factor by more specific mechanisms. Some papers are devoted to this
problem. The deterministic case of two families is discussed in
\cite{bib:determin}, and Verhulst factors are completely removed in
\cite{bib:constant}. 

\begin{figure}
\centerline{\psfig{figure=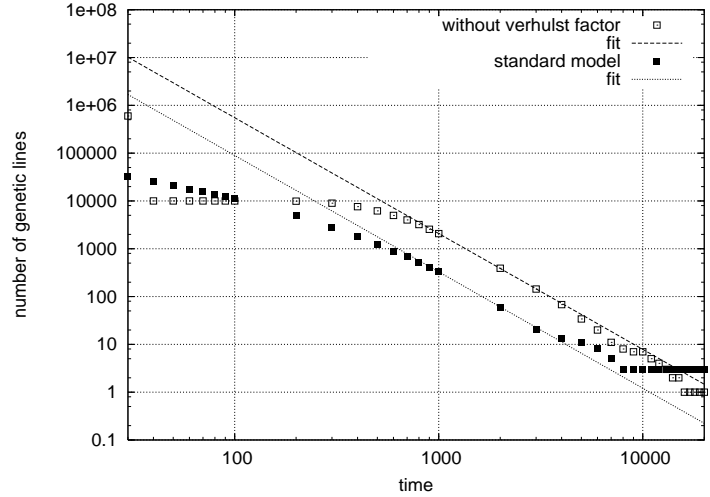,width=0.8\textwidth}}
\caption{Number of genetic families in environment as a function of time. Full
squares -- standard Penna model, empty squares -- Penna model without Verhulst
factor}
\label{fig:glines2}
\end{figure}

We explore modification introduced in \cite{bib:newverh} -- a dynamic version of
birth rate which helps to keep the size of the population stable and finite. The
birth rate changes in each time-step according to rule:

\begin{equation}
    B(t) = \left\{ \begin{array}{lcc}
    B_c \cdot (1 - \frac{n(t)}{N}) & \quad \mathrm{for} & \quad  n(t) \leq N \\
    0 & \quad \mathrm{for} & \quad n(t) > N \\
    \end{array} \right.
\end{equation}
where $B_c$ is constant (set at the beginning of simulation).  The next change
is an elimination of the Verhulst factor as a possible cause of death -- individuals
death occurs only because of genetic reasons. A comparison of results of the two
simulation -- standard and modified Penna model -- is presented in figure
\ref{fig:glines2}

Simulations were done for the same set of parameters as presented before, except
$N=610000$ for the modified model (matched to get comparable results) and
$B_c=1$. As can be seen, there is a \emph{plateau} at the initial time range in
$G(t)$ dependency, in the modified model. At the rest of time range --
dependencies are similar. The Eve effect is still present. It seems, that
replacing Verhulst factor by dynamic birth rate does not change this fact.

\section*{Conclusions}
The early stage simulation, up to about several hundred steps, corresponds to
the phase of already established size of the population and still far from the
stationary bad mutation distribution (at least for the assumed rate of the bad
mutations). Then the genetic death rate is not a dominant factor as yet, and so
the model follow the rules more as if the logistic scheme which ignores genetic
death factor only. That is why we may expect the $n=-1$ coefficient as predicted
by the logistic model.

When continuous pumping in the bad mutations yields a sort of saturation in the
bad mutation accumulations for simulation lasting long enough (say $1000$ steps
or so), the genetic death may prevail and we are inclined to associate the
$n=-2$ coefficient with this region.

Unavoidably, however we approach the final equilibrium with one family only.
Obviously this very final stage is $n=0$ and so we may anticipate a smooth
transition from $n-2$ to $n=0$ when the number of families drop down to small
number of a few competing families.

The one only final family confirms the known effect of marginal stability of
different families, when competing in a limited environment in the logistic
model. Once the genetic death is switched on (like in the Penna-like model), the
competition leads to elimination of the weaker groups. The time scale of this
process is of order of $10^5$ steps, so we need to carry on the many simulation
steps before finally the Eve emerges as the only representative family of the
whole population.

The last section on modified Penna model proves that the value of the exponent
$n$ in the power law $G(t) \propto t^n$ is indeed model-dependent. In short, the
Penna model with birth rate controlled by the current size of the population
leads to plateau $n=0$, asexual Penna model yields $n=-1$, at the beginning
several hundred of steps.

\section*{Acknowledgments}
Special thanks are due to Prof. Dietrich Stauffer. His continuous interest in
the quality and reliability of the results claimed by us, led him to significant
contributions towards our findings; he also did the simulation himself and
generously offered the results to us. Figure \ref{fig:bigsim} was produced using
his result (upper and lower curves) as well as ours (the middle lines).

We would like to express our thanks to him.

\end{document}